\documentclass[twocolumn]{aastex62}  

\usepackage{amsmath}
\usepackage{amssymb}
\usepackage{graphicx}
\usepackage{gensymb}
\usepackage{hyperref}

\makeatletter
\long\def\frontmatter@title@above{}
\makeatother

\begin{document}

\newlength{\fullw}
\setlength{\fullw}{0.5\textwidth}

\newlength{\onefigw}
\setlength{\onefigw}{0.9\fullw}

\newlength{\onefigh}
\setlength{\onefigh}{0.7\onefigw}

\newcommand{\meter}{\mathrm{m}}
\newcommand{\mm}{\mathrm{mm}}
\newcommand{\cm}{\mathrm{cm}}
\newcommand{\Mpc}{\mathrm{Mpc}}
\newcommand{\GHz}{\mathrm{GHz}}
\newcommand{\Celcius}{\mathrm{C}}

\newcommand{\aapeVa}{\texttt{PSZ2\,G156.24+22.32}}
\newcommand{\aapeVaX}{\texttt{1RXS\,J064506.9+592603}}

\newcommand{\aapeIIIa}{\texttt{PSZ2\,G138.32-39.82}}
\newcommand{\aapeIIIabis}{\texttt{RX\,J0142.0+2131}}
\newcommand{\zaapeIIIa}{0.280}

\newcommand{\aapeIIIb}{\texttt{PSZ2\,G164.18-38.88}}
\newcommand{\aapeIIIbbis}{\texttt{Abel\,401}}
\newcommand{\zaapeIIIb}{0.0739}

\newcommand{\aapeIIIc}{\texttt{PSZ2\,G144.83+25.11}}
\newcommand{\aapeIIIcbis}{\texttt{MACS\,J0647.7+7015}}
\newcommand{\zaapeIIIc}{0.584}

\newcommand{\aapeIIId}{\texttt{PSZ2\,G149.75+34.68}}
\newcommand{\aapeIIIdbis}{\texttt{RXC\,J0830.9+6551}}
\newcommand{\zaapeIIId}{0.1818}

\newcommand{\aapeIIIe}{\texttt{PSZ2\,G157.32-26.77}}
\newcommand{\aapeIIIebis}{\texttt{MACS\,J0308.9+2645}}
\newcommand{\zaapeIIIe}{0.356}

\newcommand{\planck}{\textit{Planck}}

\newcommand{\messier}[1]{\texttt{M#1}}
\newcommand{\mmfcat}{\texttt{MMF3}}
\newcommand{\pszcat}{\texttt{PSZ2}}
\newcommand{\urat}{\texttt{USNO-URAT1}}
\newcommand{\usnob}{\texttt{USNO-B1}}
\newcommand{\redmapper}{\texttt{redMaPPer}}
\newcommand{\photoobj}{\texttt{PhotoObj}}

\newcommand{\reosc}{\textsc{Reosc}}
\newcommand{\sensor}{\textsc{KAF-16803}}
\newcommand{\camera}{\textsc{Apogee U16M}}
\newcommand{\filter}{\textsc{Astrodon Gen 2 Sloan}}
\newcommand{\gcamera}{\textsc{Lodestar}}
\newcommand{\gsensor}{\textsc{Exview}}

\newcommand{\dsnine}{\textsc{DS9}}
\newcommand{\iraf}{\textsc{IRAF}}
\newcommand{\minmax}{\texttt{minmax}}
\newcommand{\avsigclip}{\texttt{avsigclip}}
\newcommand{\flatcombine}{\texttt{flatcombine}}
\newcommand{\ccmap}{\texttt{ccmap}}
\newcommand{\astutil}{\texttt{astutil}}
\newcommand{\fitparams}{\texttt{fitparams}}

\newcommand{\sextractor}{\textsc{SExtractor}}
\newcommand{\swarp}{\textsc{SWarp}}
\newcommand{\scamp}{\textsc{Scamp}}
\newcommand{\astromatic}{\textsc{AstrOmatic}}
\newcommand{\key}[1]{\texttt{#1}}

\newcommand{\classstar}{\texttt{CLASS\_STAR}}
\newcommand{\magauto}{\texttt{mag\_auto}}
\newcommand{\modelmag}{\texttt{modelMag}}
\newcommand{\cmodelmag}{\texttt{CmodelMag}}
\newcommand{\qneural}{\texttt{Q\_NEURAL}}

\newcommand{\skynet}{\textsc{SkyNet}}
\newcommand{\ash}{\texttt{ash}}
\newcommand{\R}{\textsc{R-project}}

\newcommand{\um}{\mathrm{m}}
\newcommand{\us}{\mathrm{s}}
\newcommand{\uSZ}{\mathrm{SZ}}
\newcommand{\uphot}{\mathrm{phot}}
\newcommand{\uspec}{\mathrm{spec}}
\newcommand{\uX}{\mathrm{x}}

\newcommand{\OmegaM}{\Omega_\um}
\newcommand{\zphot}{z_\uphot}
\newcommand{\zspec}{z_\uspec}
\newcommand{\zx}{z_\uX}
\newcommand{\Msz}{M_{\uSZ}}
\newcommand{\Mf}{M_{500}}
\newcommand{\Msun}{M_{\sun}}
\newcommand{\ts}{\theta_\us}
\newcommand{\YfR}{Y_{\mathrm{5R500}}}

\title{\large\textbf{Optical Follow-up of Planck Cluster Candidates
    with Small Instruments}}

\author{Vincent Boucher}
\affiliation{CREEF asbl, 3 rue d'Ostin, 5080 La Bruy\`ere, Belgium}
\author{Simon de Visscher}
\affiliation{4 rue du coll\`ege, 1407 M\'ezery-pr\`es-Donneloye,
  Switzerland}
\author{Christophe Ringeval}
\affiliation{Centre for Cosmology, Particle Physics and Phenomenology,
  Institute of Mathematics and Physics,\\ Louvain University,
  2 chemin du cyclotron, 1348 Louvain-la-Neuve, Belgium}
\email{christophe.ringeval@uclouvain.be}

\date{\today}

\begin{abstract}
  We report on the search for optical counterparts of {\planck}
  Sunyaev-Zel'dovich (SZ) cluster candidates using a $0.6\,\meter$
  non-professional telescope. Among the observed sources, an
  unconfirmed candidate, {\aapeVa}, is found to be associated with a
  region of more than $100$ galaxies within a $3$ arcminutes radius
  around the Sunyaev-Zel'dovich maximum signal coordinates. Using $14$
  hours of cumulated exposure over the Sloan color filters $g'$, $r'$,
  $i'$, and $z'$, we estimate the photometric redshift of these galaxies
  at $\zphot=0.29 \pm 0.08$. Using the red-sequence galaxy method
  gives a photometric redshift of $0.30{\,}^{+0.03}_{-0.05}$. Combined
  with the {\planck} SZ proxy mass function, this would favor a
  cluster of $4.4 \times 10^{14}$ solar masses. This result suggests
  that a dedicated pool of observatories equipped with such
  instruments could collectively contribute to optical follow-up
  programs of massive cluster candidates at moderate redshifts.
\end{abstract}


\section{Introduction}

Among the foregrounds imprinting temperature anisotropies and spectral
distortions on the Cosmic Microwave Background (CMB) radiation, the
thermal Sunyaev-Zel'dovich effect is of particular interest for
Astrophysics and Cosmology~\citep{Sunyaev:1970, Sunyaev:1972,
  Sunyaev:1980, Sazonov:1999zp, Challinor:1999yz,
  Bunn:2006mp}. Inverse Compton scattering of the CMB photons by the
hot electrons present within galaxy clusters results in a shift and
distortion of their blackbody spectrum towards higher frequencies. The
intracluster plasma cools down the CMB radiation at frequencies
typically lower than $217\,\GHz$ while warming it up at higher
frequencies. Such a signature is unique among foregrounds and has been
intensively used by the {\planck} satellite collaboration and other
CMB ground telescopes, such as the Atacama Cosmology Telescope (ACT)
and the South Pole Telescope, to provide unprecedented catalogues of
Sunyaev-Zel'dovich (SZ) sources~\citep{Hasselfield:2013wf,
  Ade:2013skr, Bleem:2014iim, Ade:2015mva}. The SZ effect does not
depend on the redshift, and therefore provides a new high-redshift
observable of galaxy clusters. The full sky coverage of the {\planck}
satellite has allowed the release of the {\pszcat} catalog in
\citet{Ade:2015gva} containing more than $1600$ SZ sources. Among them,
$1200$ objects have been confirmed as clusters by the {\planck}
collaboration through their cross identification in the Meta-Catalog
of X-ray detected clusters (MCXC)~\citep{2011A&A...534A.109P}, with
optical counterparts in the Sloan Digital Sky Survey
(SDSS)~\citep{York:2000gk}, in the {\redmapper}
catalog~\citep{Rozo:2014zva, Rykoff:2016trm}, in the Nasa/IPAC
Extragalactic Database (NED), and with infrared galaxy overdensities
in the Wide-field Infrared Survey
(WISE)~\citep{2010AJ....140.1868W}. More than $400$ SZ sources were
still unconfirmed at the time of publication of the catalog and not
all confirmed sources have redshift information. Cluster counts using
the {\pszcat} catalogue have been used for Cosmology in
\citet{Ade:2015fva}. They allow CMB-independent estimations of the
amplitude of the matter power spectrum $\sigma_8$ and of the matter
density parameter $\OmegaM$, which are currently in mild tension with
the best fit $\Lambda$CDM model obtained from CMB primary anistropies
alone~\citep{Ade:2015xua}. Doing cosmology with cluster counts
requires the determination of a scaling relation between the total
integrated Compton parameter $\YfR$, the cluster angular size, $\ts$,
and its total mass, $\Mf$, (defined over a radius enclosing $500$ times
the critical density at redshift $z$). As explained in
\citet{Ade:2013lmv}, SZ measurements give information on the relation
between $\YfR$ and $\ts$ while breaking the degeneracy between $\YfR$
and $\Mf$, at a given redshift $z$, relies on a scaling relation
extracted from X-ray observations and assuming hydrostatic equilibrium
of the intracluster gas~\citep{2010A&A...517A..92A,
  Arnaud:2007br}. For each of the SZ sources, the {\pszcat} catalog
provides the most probable hydrostatic mass $\Msz(z)$ (and its
standard deviation) assuming the scaling relation to hold. Follow-up
programs of SZ sources are therefore of immediate interest in
evaluating $\Msz$ for a given cluster by the determination of its
redshift $z$.

Dedicated optical follow-up programs of unconfirmed {\planck} clusters
have been carried on using professional telescopes since the
publication of the {\pszcat} catalog. The Pan-STARRS $1.8\,\meter$
telescope survey~\citep{Liu:2014rza} has provided $60$ confirmations
combined with spectroscopic redshift measurements.  Another $16$
high-redshift clusters have been confirmed by the
Canada-France-Hawa\"{\i} telescope (CFHT) with photometric redshifts
together with richness mass
estimates~\citep{vanderBurg:2015ssn}. Spectroscopic redshifts and
confirmation of $13$ more clusters have been provided by the
$1.5\,\meter$ Russian-Turkish telescope, the $2.2\,\meter$ Calar Alto
Observatory telescope, and the $6\,\meter$ Bolshoi (BTA) telescope in
\citet{Vorobyev:2016hez} \citep[see also][]{Ade:2014pna}. The
{\planck} collaboration has used a month of observations at the
Canary Islands Observatories for the confirmation of $73$ more
candidates~\citep{Ade:2015qqa}. Although these numbers could suggest
that unconfirmed SZ sources would soon be exhausted, artificial neural
networks have recently been used on the {\planck} CMB measurements to
improve the detection threshold of SZ sources. A new catalog discussed
by \citet{Hurier:2017krx} contains almost $4000$ galaxy cluster
candidates. In addition, ground-based telescopes are still
contributing to new SZ detection thereby calling for an increase of
the telescope time needed for follow-up
programs~\citep{Hilton:2017gal} .

In this paper, we explore and report on the possibility to use
sub-meter non-professional telescopes (of typically
$0.5\,\meter$-$0.6\,\meter$ diameter) to carry on optical follow-up
searches of unconfirmed SZ sources. In spite of obvious technical
challenges, some scientific measurements can be achieved such as
galaxy counts and photometric redshift estimates. As a test case, we
report on a cluster candidate, {\aapeVa}, which has not been confirmed
in any of the aforementioned catalogs, nor in the follow-up programs,
and which is not included in the SDSS sky
coverage~\citep{2012ApJS..203...21A, 2012ApJS..199...34W}.

The paper is organized as follows. Section~\ref{sec:obs} describes the
observations and the method used for data reduction. Some details are
provided concerning the observatory, the hardware used, as well as the
technical difficulties overcome for achieving sufficiently good
tracking and imaging for photometry. In Sect.~\ref{sec:results}, we
present our main results, namely galaxy counts, photometric redshift,
and optical richness estimates for {\aapeVa}. In
Sect.~\ref{sec:discuss}, various tests to assess the validity of our
measurements are presented. In the conclusion, we briefly discuss a
proof of concept on what small telescopes can and cannot achieve,
would they massively contribute to the search of optical counterparts
of SZ sources.

\section{Observations and data reduction}
\label{sec:obs}

\subsection{Observatory and instrumentation}

The observatory of Saint-V\'{e}ran is located in France, at
$44\degree 42\arcmin 03\arcsec$ N, $6\degree 52\arcmin 06\arcsec$ E, at an 
altitude of
$2930\,\meter$ a.s.l. (IAU code 615). It has been managed by the
Astroqueyras Association\footnote{\url{https://www.astroqueyras.com}}
for roughly $30$ years, and is intended to allow amateur astronomers
to observe under better conditions compared to that of typical lower
altitude suburban skies.

The observatory comprises three main instruments: a Cassegrain
telescope with a diameter of $0.62\,\meter$ and $9\,\meter$ of focal
length (T62), and two Ritchey-Chr\'{e}tien telescopes, each having a
diameter of $0.5\,\meter$ and $4\,\meter$ of focal length. The
instrument mainly used in the context of this work is the
T62. Although the primary mirror is of $0.62\,\um$ diameter, the
holding baffle reduces the effective aperture to $0.60\,\um$. The T62
is mounted on a German-type {\reosc} equatorial mount, equipped
electronically for semi-automatic target pointing. The imaging device
used is an {\camera} CCD camera equipped with a {\sensor} sensor,
cooled at $-20\degree\,\Celcius$. The sensor has $4096 \times 4096$
pixels of area $9\times9\,\micron^2$, which implies a genuine angular
sampling of $0.21\arcsec \times 0.21\arcsec$. The field of view is a
square of $14\arcmin$ side.

The \citet{1989S&T....77S.387K} polar drift method has been applied to
align the mount right ascension (R.A.) and Earth rotation axes, and
therefore improves the pointing and tracking accuracy. This operation
must be done regularly, due to local seismic activity and a slow
rotation of the mountain where the observatory is installed. In the
context of this work, the residual angle between the right ascension
axis of the mount and the Earth rotation axis was reduced to a value
smaller than roughly $10\arcsec$. The native periodical error of the
RA movement is estimated to be around $5\arcsec$ at vanishing
declination (DEC), with smooth amplitude variations during
tracking. The tracking accuracy is yet improved by auto-guiding the
T62, i.e., by correcting any deviations between the sky and mount
movements. This is achieved using a second camera installed on a
$0.2\,\meter$ Schmidt-Cassegrain telescope. It is mounted in parallel
with the T62 and has a focal length of $2\,\meter$. The camera has a
{\gsensor} sensor made of $752 \times 580$ pixels of area $8.2 \times
8.4\,\micron^2$. This allows for exposures up to $10$ to $15$ minutes
long, depending on the telescope orientation and the target
declination.  The resulting guiding error is estimated from the
average point spread function of the stars and found to be less or
equal to the seeing. The measured full width at half maximum (FWHM)
varies between $1.2\arcsec$ to $2\arcsec$ across the exposure series.
These FWHM values led us to optimize the sampling for the {\camera}
camera. A binning grouping four pixels into one has been used,
yielding an angular sampling of $0.42\arcsec \times 0.42\arcsec$.

The search for optical counterparts of the SZ targets has been
performed in luminance, i.e. without a specific spectral filtering of
the light input. Once the galaxies have been located, color imaging
of the suspected cluster is done using interferential filters. For
this work, we have used a set of {\filter} photometric filters
corresponding to the SDSS photometric
standards~\citep{Fukugita:1996qt, 2002AJ....123.2121S}. The low
quantum efficiency of the {\sensor} sensor in the ultraviolet
wavelength range led us to consider only the $g'$, $r'$, $i'$, and $z'$
filters.

\subsection{Data acquisition}
\label{sec:acquisition}

The search of optical counterparts of the {\planck} SZ candidates took
place during two one-week long observation missions, one from 2015
November 7 to November 15, and the other between 2017 February 18 and
February 26.

The goal of the first mission was to assess the feasibility of the
project and was dedicated to the observations of confirmed {\planck}
clusters having known redshifts. Individual galaxy identifications for
five clusters having redshifts ranging from $0.1$ to $0.6$ were
successfully performed. In particular, color imaging of two of them,
{\aapeIIIa} and {\aapeIIIb}, located at redshift $0.280$ and $0.08$,
respectively, allowed us to test the instrumentation for photometric
redshift estimation. A low-signal test of our apparatus was achieved
by successfully resolving individual galaxies in luminance within a
distant cluster {\aapeIIIc}, having a redshift of $0.584$ (also known
as {\aapeIIIcbis}). These results are discussed in more details in
Sect.~\ref{sec:discuss}.

The 2017 winter mission was dedicated to the search of non-confirmed
{\planck} SZ sources optical counterparts. Five objects from the
{\pszcat} catalog, having no X-ray nor optical counterparts, have
been pointed. The choice of the targets has been made taking into
account several constraints. First, we have kept the airmass
reasonably close to unity and only objects well-above the horizon, by
at least $40\degree$, have been considered. Second, the targets should
not be too close from the Milky Way to minimize possible confusion
between faint stars and distant galaxies.

A technical difficulty associated with the use of a small
telescope is the exposure time needed to collect enough photons for
galaxy detection. Based on empirical results from the first mission,
around $1$ hour of exposure has been accumulated per target, made from
$10$ minutes luminance slices. After performing some minimal data
reduction, a visual inspection allowed us to check for overdensity
within the $14\arcmin$ field of view around the SZ signal coordinates.

Among the observed targets, {\aapeVa} located in the Lynx
constellation (northern celestial hemisphere), ended up showing a
significant object count. We therefore started data acquisition in the
color filter $g'$, $r'$, $i'$ and $z'$ from 2017 February 21 until
the end of the mission on February 26. According to the
above-mentioned criteria, {\aapeVa} was of sufficiently high altitude
in the sky only during the second half of the night. The first part of
the night was therefore dedicated in the color imaging of our
reference photometric SDSS DR9 galaxy field, located nearby
{\messier{63}}, which is roughly at the same altitude in the sky than
the one occupied by the cluster later on. Around $1.5$ hours of
cumulated exposure per filter were taken for the reference field. In
total, around $14$ hours of color exposure have been used to derive
the results presented below.

\subsection{Reduction}
\label{sec:reduc}

In order to produce photometric usable data, image reduction has been
performed in two steps.

In a first pass, we have used the Image Reduction and Analysis
Facility ({\iraf}) to correct each image from bias, dark currents and
nonlinear response pixels~\citep{1986SPIE..627..733T,
  1993ASPC...52..173T}. Basic astrometric registration and airmass
calculations are also performed for each color images in order to
provide a World Coordinate System (WCS) compatible header for each of
them.

\begin{figure}
\begin{center}
  \includegraphics[width=\onefigw]{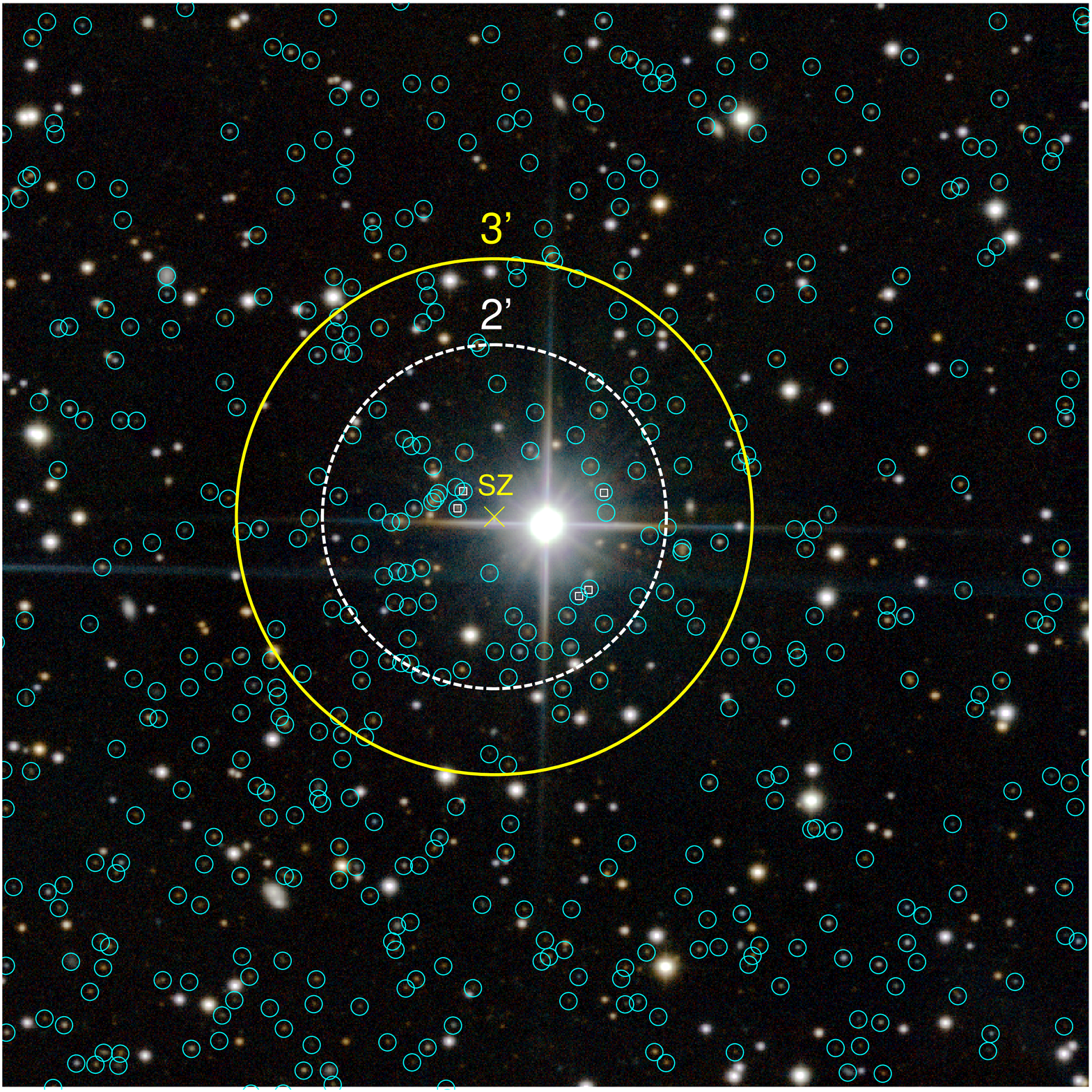}
  \includegraphics[width=\onefigw]{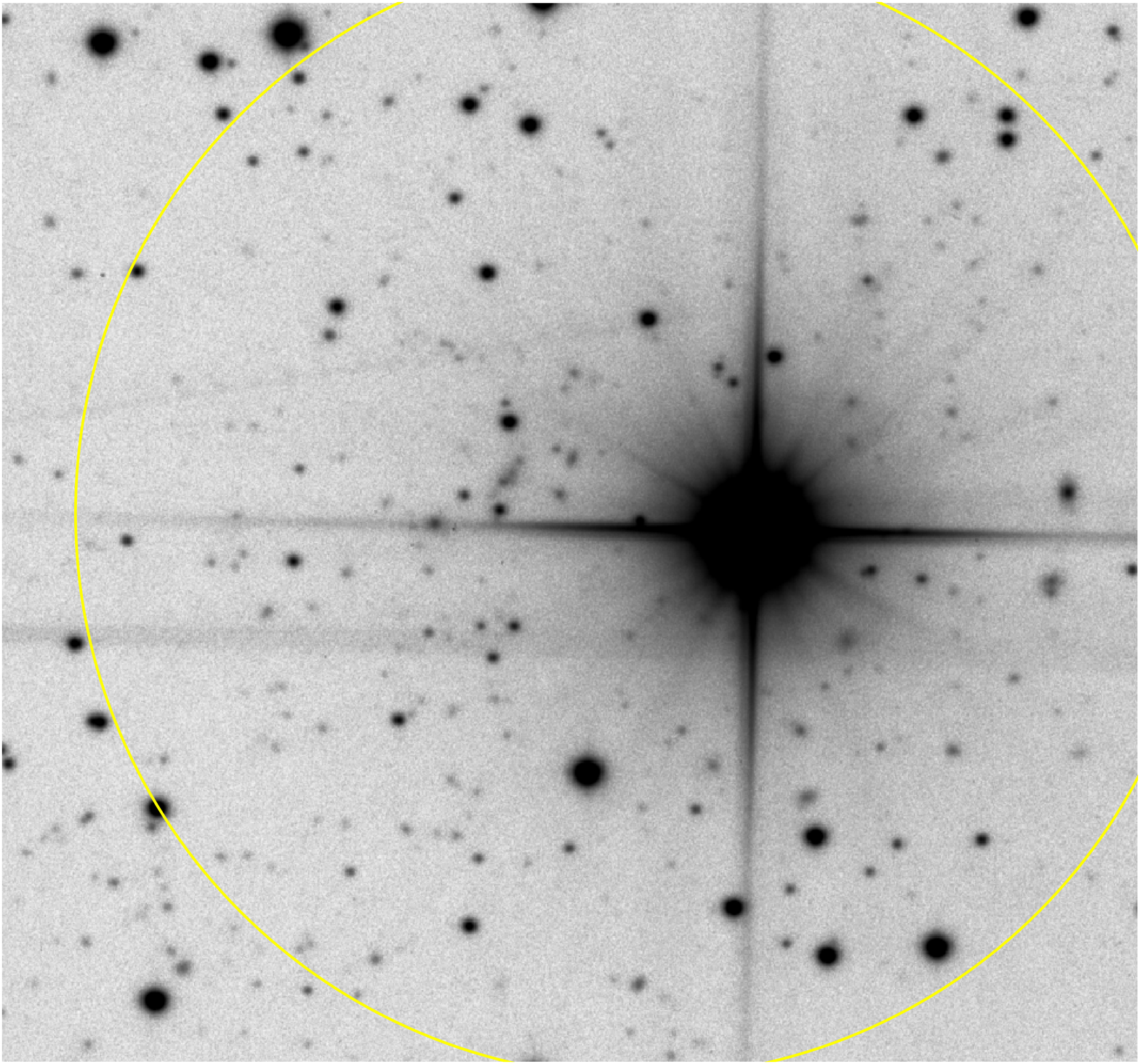}
  \caption{Upper panel: circular regions of radius $2\arcmin$ and
    $3\arcmin$ centered at the {\planck} SZ maximum signal location
    (marked by a cross) used for galaxy counts and photometric
    redshift estimates. Objects with a small circle are matched
    sources across the color filters (stars excluded; see the text). The
    lower panel is a zoom within the $3\arcmin$ region.}
\label{fig:regions}
\end{center}
\end{figure}

The second pass consists in an accurate astrometry and photometry
solving starting from the WCS single-epoch images produced by the
first pass. For this purpose, we have used the {\astromatic} software
suite~\citep{1996A&AS..117..393B, 2006ASPC..351..112B,
  2002ASPC..281..228B}. Sources are extracted in all images using
{\sextractor} (version 2.19.5) with a detection and analysis threshold
set at $1.5\sigma$ with respect to the local background noise. An
astrometric and photometric solution is then obtained by using the
{\scamp} code. It has been run in multi-instruments mode, one
astrometric and photometric instrument for each color filter, plus
luminance, and we have allowed for astrometric distortions within a given
instrument. The reference astrometry catalog chosen is the
{\usnob}~\citep{2003AJ....125..984M}. The reported internal
astrometric errors are small, $0.05\arcsec$ (FWHM), and remain
negligible with respect to the ones of the reference catalog,
evaluated at $0.5\arcsec$ (FWHM). However, a shear contraction
amplitude of typically $0.15\%$ has been corrected for each
instrument, far above what one could have expected for an airmass
never exceeding $1.4$ ($0.05\%$). The distortion map of the pixel
scale computed by {\scamp} indeed reveals an almost constant
hyperbolic pattern over all filters and luminance. Although the reason
for this distortion is unclear, different sources are possible, among
which a stress on the primary mirror, or still a minor mirror
alignment issue leading to astigmatism appearing at moderate
defocus. Time constraint did not allow for a deeper exploration and
resolution of this issue. Concerning the photometric solution,
zero-point changes across the exposure series have been corrected
within each filter. The standard deviations of these corrections are
found in the following ranges: $\Delta g'_0 \in [0.1,0.4]$, $\Delta
r'_0 \in [0.05,0.3]$, $\Delta i'_0 \in[0.05,0.4]$ and $\Delta z'_0
\in[0.05,0.5]$. The first (and lowest) value is for high
signal-over-noise matched sources only while the second one includes
all matched objects between the single-epoch images and the reference
catalog. Although corrected, these numbers show some degradation of
our photometric stability up to half a magnitude for faint objects in
the $z'$ filter, and some significant scatter in the $g'$ band as
well. This could be attributed to the parasitic presence of a $V=7.2$
magnitude double star in the field of view as a much lower scatter is
obtained for the SDSS reference field (see below and
Fig.~\ref{fig:regions}).

In the last step, coaddition of the corrected single-epoch images for
each filter (and luminance) has been delegated to the {\swarp} code,
which takes care of background estimation, as well as resampling, for
each image. Highest contrasts have been obtained by using a median
weighted stacking, the weights being estimated from the
background noise~\citep[see][]{2002ASPC..281..228B}.

\subsection{Calibration}

The absolute zero-point magnitudes for the four coadded images in the
$g'$, $r'$, $i'$, and $z'$ filter have been determined from our SDSS DR9
reference galaxy field (see Sect.~\ref{sec:acquisition}).

The reference images have been reduced and flattened, for each color
filter, exactly as described above. Then, we have used the
{\sextractor} code over the coadded images to create a catalog of
photometric reference sources. The magnitude measurement method chosen
is {\magauto} and only high signal-over-noise detections have been
kept, $5\sigma$ above the estimated
background~\citep{1996A&AS..117..393B}. Our reference catalogs, in
$g'$, $r'$, $i'$, and $z'$, are then matched together and to the
SDSS DR9 catalog by cross identifying objects of same position at
less than $1\arcsec$. For this purpose, the imaging software {\dsnine}
has been used to create a set of roughly $30$ galaxies for which one
has both the non-calibrated {\magauto} magnitudes and the {\modelmag}
magnitudes as provided by the SDSS DR9
catalog~\citep{2012ApJS..203...21A}. The zero-point values for each
filter, $g'$, $r'$, $i'$ and $z'$, have been determined with {\iraf}
using a photometric weighting based on the magnitude error estimates
made by {\sextractor}.

In the following, unless specified otherwise, our calibrated
{\magauto} magnitudes will be referred to as $griz$ since they are
indistinguishable of the {\modelmag} SDSS DR9 magnitudes within the
uncertainties of our apparatus and methodology.

\section{Results}
\label{sec:results}

\subsection{Color catalog of matched objects}
\label{sec:colorcat}

From the photometric usable images obtained as described in
Sect.~\ref{sec:obs}, we have constructed a $griz$ catalog of all
objects suspected to be galaxies within the $14\arcmin$ field of view
around the SZ coordinates associated with {\aapeVa}.

Again, the {\sextractor} code has been used for this purpose, in
single-image mode and with a detection threshold at $2\sigma$ for the
$g'$, $r'$, and $i'$ filter. As discussed in the previous section, the
$z'$ image showing higher noise than the others, {\sextractor} has
been used in two-images mode, $2\sigma$ detection coming from the $i'$
image, photometry coming from the $z'$ one with an analysis threshold
set at $1.5\sigma$. Separation between stars and galaxies is made
using the neural network classifier of {\sextractor} by discarding all
objects having a {\classstar} greater than $0.98$. The remaining objects
over all filters are then matched together according to their
position, required to be the same at $1\arcsec$, and then rejected if
they correspond to an optical counterpart in the {\urat}
catalog~\citep{2015AJ....150..101Z}.

In Fig.~\ref{fig:regions}, the SZ signal center, as reported in the
{\pszcat} catalog, has been represented by a cross and small circles
have been drawn around all the remaining objects of our $griz$
catalog. A visual inspection of their shape suggests that many of them
should be galaxies while some overdensity is visible left from the SZ
center. Let us notice that having chosen to exclude objects from the
{\urat} catalog is unexpectedly doing a good job in removing bright
and close galaxies visible in the picture. A closer look to
Fig.~\ref{fig:regions} also reveals that many small objects close to
the bright star, and which are certainly galaxies belonging to the
cluster, are not present in our $griz$ catalog. This is due to the
fact that these sources have been excluded by {\sextractor} as being
too much contaminated, in at least one color, by the parasitic glow of
the star. As discussed below, five more objects close to the star end
up being obvious outliers and will be removed from the subsequent
analysis (drawn with a small white box).

From the {\planck} catalog {\mmfcat}, we have extracted the
two-dimensional probability distribution in the plane $(\YfR,\ts)$
associated with {\aapeVa} \citep{Ade:2015gva}. By marginalizing over
the integrated Comptonization parameter $\YfR$, the most probable
angular size of the SZ region is found to be
\begin{equation}
  \ts = 2.5\arcmin \pm 0.5\arcmin,
\label{eq:ts}
\end{equation}
at $68\%$ of confidence. As a result, we have defined a circular
region around the SZ maximum having a radius of $3\arcmin$ to
encompass the whole SZ region. It is represented as the yellow circle
in Fig.~\ref{fig:regions}. To assess the sensitivity of the results
with respect to the chosen cluster angular size, another circular
region has been defined, centered at the SZ location, with a radius of
$2\arcmin$ (dashed inner circle in Fig.~\ref{fig:regions}). The
intersection of our catalog with these two regions will be referred to
as $griz_3$ and $griz_2$ in the following; they contain $114$ and $58$
objects, respectively (including the five outliers; see
Fig.~\ref{fig:regions}).

Finally, let us mention that {\aapeVa} is associated in the {\pszcat}
catalog with a value of $\qneural=0.96$, which indicates a high
probability of being a genuine SZ source. Moreover, as reported by
\citet{vanderBurg:2015ssn}, all the SZ candidates validated by optical
counterparts in this work have a $\qneural>0.8$, thereby rendering our
observation of galaxy overdensity around the SZ location not unlikely.

\subsection{Photometric redshift and cluster mass}

\begin{figure}
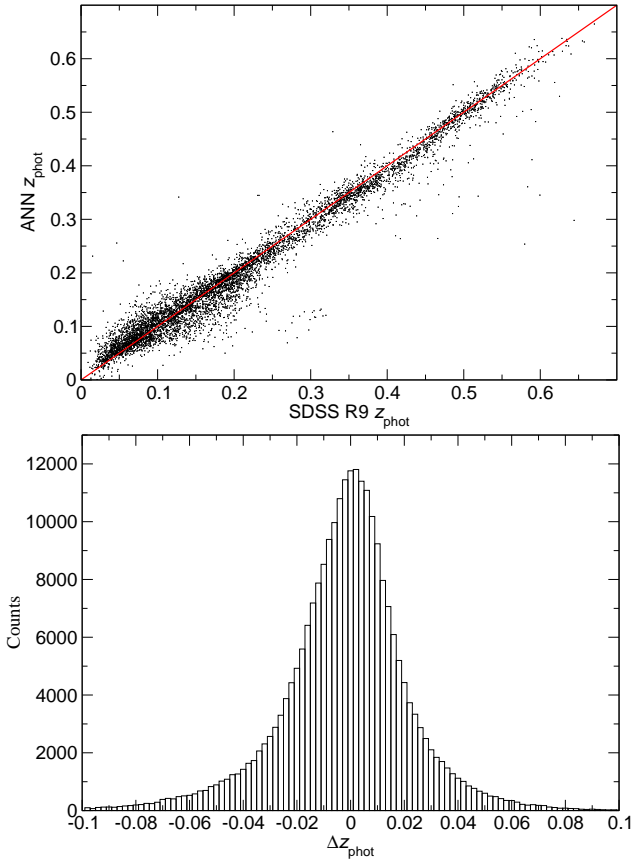

  \begin{center}
    \includegraphics[height=\onefigh]{figures/zann_test_pred_4x90_2}
    \includegraphics[height=\onefigh]{figures/zann_errors}
    \caption{Photometric redshift estimates of the artificial neural
      network over an independent testing set of $250\,000$ galaxies
      from the SDSS DR9 catalog. The upper panel shows the neural net
      estimated redshift $\zphot$ as a function of photometric SDSS
      redshift (the plot has been truncated to $8000$ random samples
      for illustration convenience). The lower panel shows the
      distribution of the residuals $\Delta \zphot$ and has a standard
      deviation of $\sigma_{\Delta \zphot} = 0.027$.}
\label{fig:anntest}
  \end{center}
\end{figure}

\begin{figure}
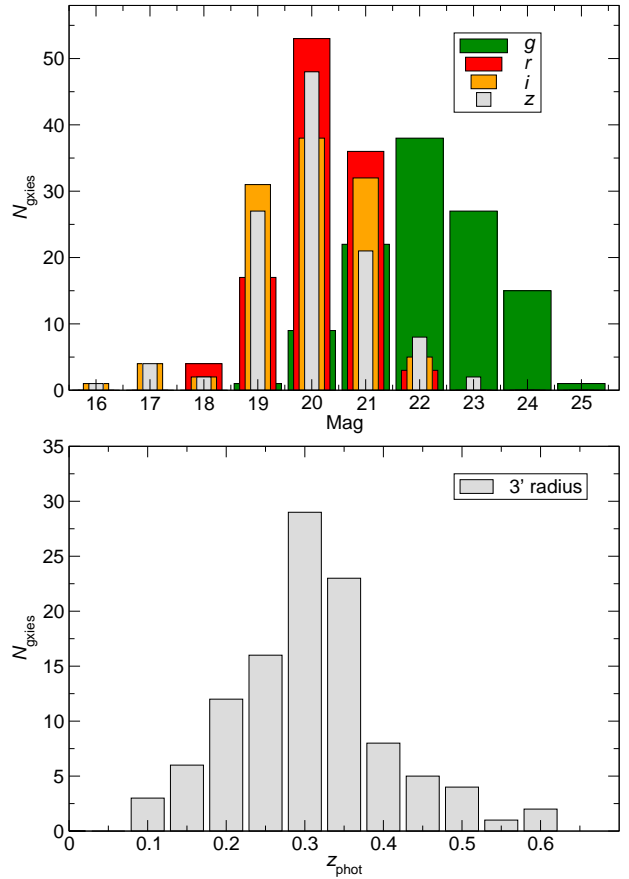

\begin{center}
  \includegraphics[width=\onefigw]{figures/mag_auto_3arcmin}
  \includegraphics[width=\onefigw]{figures/zphot_3arcmin}
  \caption{Upper panel: magnitude distribution in $griz$ for the $109$
    galaxies observed within the $3\arcmin$ radius region centered at the SZ
    coordinates. The lower panel is the corresponding photometric
    redshift distribution as inferred by our neural network trained
    over the SDSS DR9 catalog (five outliers having negative redshifts
    removed).}
\label{fig:griz3}
\end{center}
\end{figure}

In order to evaluate the photometric redshift of the objects belonging
to our color catalogs, we have used an artificial neural network
(ANN) regression of the SDSS DR9 color-redshift relationship, but only
for the reduced set of $griz$ colors. For this purpose, the publicly
available {\skynet} code has been run to train a simple feed-forward
artificial neural network~\citep{2012MNRAS.421..169G,
  2014MNRAS.441.1741G, 2014IAUS..306..279H}. The ANN input layer takes
as argument the $griz$ {\modelmag} magnitudes, and the output layer
returns the photometric redshift. All input and hidden layer nodes of
the ANN are evaluating the hyperbolic tangent of the biased weighted
addition of the input signals. For faster training, the output layer
has, however, been chosen to only perform a biased linear combination of
the last hidden layer nodes. The training set chosen consists of
$250\,000$ photometric clean objects from the SDSS DR9 data release,
selected to have nearest neighbors in the SDSS color space,
photometric redshift errors not exceeding $0.03$, and color magnitudes
in the range $[12,29]$. Another group of $250\,000$ objects, selected
with the same criteria, has been used as a testing set to check the
quality of the ANN regression. We have found that a good enough
learning of the ANN over the training set requires four hidden layers
of $90$ nodes each, in addition to the input layer of four nodes and
the single-node output layer. In total, {\skynet} has been used to
find the optimal values of more than $30\,000$ weights and biases as
described in~\citet{2014MNRAS.441.1741G}.

The upper panel of Fig.~\ref{fig:anntest} shows the ANN output values
obtained by inputting the $griz$ color magnitudes of the SDSS DR9
testing set as a function of the photometric redshift estimated by the
SDSS DR9 collaboration. The lower panel shows the distribution of the
residuals. The ANN reproduces very well the SDSS estimated photometric
redshifts for almost all values from $\zphot \simeq 0$ to $\zphot
\simeq 0.8$. The standard deviation of the residuals, over a quarter
million of objects, reads $\sigma_{\Delta\zphot} = 0.027$. Because it
is of the same order as the SDSS photometric redshift errors of the
training catalog, the ANN regression is nearly optimal.

Applying the trained ANN over the color magnitudes of our $griz_3$
catalog gives the distribution of photometric redshifts of all
galaxies located at less than $3\arcmin$ from the SZ center associated
with {\aapeVa}. In total, $griz_3$ contains $114$ objects, five of them
end up having a negative photometric redshift. A closer examination of
these outliers show that they have unusually large values for the
estimated error on {\magauto} ($\Delta m \gtrsim 0.2$), and they are
all located close to the bright star. They have therefore been removed
from the analysis and are represented as white boxes in
Fig.~\ref{fig:regions}. The photometric redshift distribution of all
the $109$ remaining objects has been plotted in the lower panel of
Fig.~\ref{fig:griz3} while the upper panel shows their color
magnitude distribution. A clear reddening of the objects can be
observed between the $g$ and $r$ magnitude, and to a lesser extend
between $r$ and $i$ as well, supporting the claim that the sources are
redshifted galaxies. The photometric redshift distribution indeed
shows a peak around a redshift of $0.3$ and no object are found with a
redshift less than $0.08$ (up to the five outliers). In order to
estimate the mode of the redshift distribution, the ``{\ash}'' package
of the {\R} software suite has been used to compute a polynomial
density estimate and extract its maximum. One finds
\begin{equation}
  \zphot=0.29 \pm 0.08.
\label{eq:zphot3}
\end{equation}
To minimize sensitivity to potential systematics, the quoted
error stands for the median absolute deviation around the mode,
normalized such that it would match the usual standard deviation for a
Gaussian distribution. Changing the smoothing kernel does not affect
the estimation of the mode by more than $0.01$. As expected, the
median absolute deviation is larger than the intrinsic ANN residuals
plotted in Fig.~\ref{fig:anntest} due to the color-magnitude errors of
our measurements.

The most probable hydrostatic mass, $\Msz(z)$, as well as its standard
deviation, as a function of the cluster redshift $z$ is given by the
{\planck} collaboration in the {\mmfcat}
catalog~\citep{Ade:2015gva}. Combined with Eq.~\eqref{eq:zphot3}, it
yields
\begin{equation}
\Msz = \left( 4.4\pm 1.3 \right)\times 10^{14} \Msun.
\label{eq:msz}
\end{equation}
The quoted errors are the intersect of the median absolute deviations
associated with the photometric redshift and the standard deviation on
$\Msz(z)$.

In order to test the robustness of Eq.~\eqref{eq:zphot3} with respect
to the chosen condition under which galaxies belong to the cluster, we
have also considered a region of $2\arcmin$ radius around the SZ
coordinates. This corresponds to the lower $1\sigma$ limit of $\ts$ in
Eq.~\eqref{eq:ts}, and to the objects constituting the $griz_2$ color
catalog. After the removal of the same five outliers as in the
$griz_3$ catalog, it remains $53$ galaxies for which one finds
$\zphot=0.29 \pm 0.11$. As before, the first value is the mode and the
error stands for the median absolute deviation around the mode.

\subsection{Optical richness}

\begin{figure}
  \begin{center}
    \includegraphics[height=0.89\onefigw]{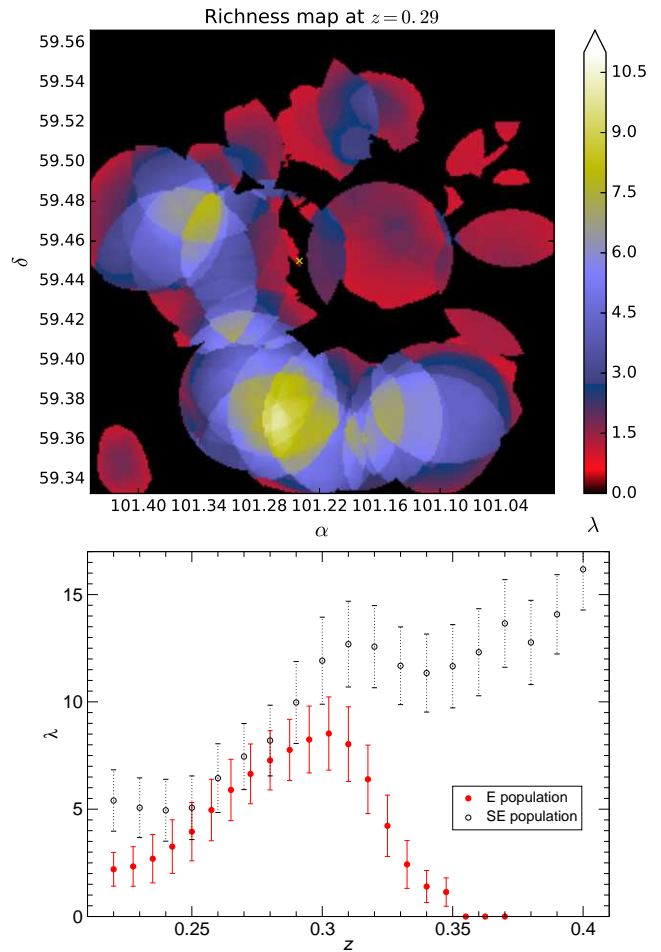}
    \includegraphics[height=0.94\onefigh]{figures/maxrichness}
    \caption{Upper panel: map of the optical richness $\lambda$ over
      the field of view of figure~\ref{fig:regions} (the red sequence
      galaxies are assumed to be at $z=0.29$). The lower panel shows
      the dependence of $\lambda(z)$ for the two locations exhibiting
      maximal richness in the upper panel. Only the one located East
      to the SZ center is well bounded in redshift and can be
      associated with a galaxy cluster (red points).}
\label{fig:richmap}
  \end{center}
\end{figure}

A visual inspection of the red luminous objects contiguous to the ones
contained within the $2\arcmin$- and $3\arcmin$-radius regions
suggests that a galaxy population is also present further southest
(SE), i.e., toward the lower left corner of
Fig.~\ref{fig:regions}. Let us notice that the NED database reports
one X-ray source in this region,
{\aapeVaX}~\citep{1999A&A...349..389V}.

To quantify which objects are clustered, we have estimated
the optical richness, $\lambda$, as discussed in
\citet{2009ApJ...703..601R}. In practice, $\lambda$ has been computed
according to the method presented in \citet{2012ApJ...746..178R} and
\citet{2014ApJ...785..104R}. In order to use the same luminosity
tracer as in these references, we have recalibrated our $i'$ filter
over the SDSS DR9 reference field by using the $i$ {\cmodelmag}
magnitude (instead of the {\modelmag}). The color has been estimated
as before, namely our $g-r$ comes from $g'$ and $r'$ filter calibrated
from {\modelmag} magnitudes. This color should be an accurate tracer
of the red-sequence galaxies up to redshift
$0.35$~\citep{2000AJ....120.2148G, 2007ApJ...660..221K}. All of the
magnitudes have been corrected for Galactic extinction and
reddening~\citep{1998ApJ...500..525S, 2011ApJ...737..103S}, and the
cosmological parameters have been chosen according to the Planck 2015
favored values~\citep{Ade:2015xua}.

The presence of the parasiting star and the other uncertainties
associated with our small instrumentation are expected to introduce
various systematics in the determination of the photometric
zero-points. As reported in \citet{2012ApJ...746..178R}, the
determination of $\lambda$ is expected to be accurate for zero-point
shifts up to $\pm 0.05$ magnitudes, and we might have systematic
uncertainties as large as eight times this value for some objects (see
section~\ref{sec:reduc}). Adding an additional scatter of $0.1$
magnitudes to luminosity and colors is indeed found to increase the
value of $\lambda$ by a factor of $2$. As a result, there is a
possibility that the following evaluations of $\lambda$ are
underestimated, which is inherent to the experimental limitations of
our setup. Notice, however, that the dependence of $\lambda$ with
respect to the coordinates of the clustering center and to the
redshift, remains mostly insensitive to the uncertainties in the
photometric zero-points. This is what we focus on in the following.

To minimize the propagation of these potential systematics,
we have chosen the radial filter to be
\begin{equation}
  R_{\mathrm{c}}(\lambda) = R_0 \left(\dfrac{\lambda}{100}\right)^{\beta},
\end{equation}
with $R_0 = 0.95 h^{-1}\,\Mpc$, and $\beta=0.1$ instead of $R_0 =1
h^{-1}\,\Mpc$ and $\beta=0.2$ as done in
\citet{2012ApJ...746..178R}. Our choice is still in the region of
minimal intrinsic scatter for red-sequence galaxies, but has the
advantage of lowering the impact of underestimated $\lambda$
values. The background galaxies and the red-sequence fit in $g-r$
have been taken as in \citet{2012ApJ...746..178R} while the luminosity
filter is the one of \citet{2014ApJ...785..104R}. 

The upper panel of figure~\ref{fig:richmap} shows a map of $\lambda$
over the field of view of figure~\ref{fig:regions}, obtained by
postulating that the red-sequence galaxies are at the redshift
$z=0.29$. The two regions exhibiting maximal richness encompass the
one visually expected east from the SZ center (left), around
$(\alpha,\delta)=(101.31,59.48)$, and the one associated with the SE
population at $(\alpha,\delta) = (101.27,59.36)$ (lower left). In the
lower panel of figure~\ref{fig:richmap}, we have plotted the optical
richness $\lambda(z)$, as a function of the redshift, for these two
locations. Only the first one ends up being well localized in redshift
space, and this allows us to improve the cluster's redshift estimate to
\begin{equation}
z_{\lambda} = 0.30{\,}^{+0.03}_{-0.05}\,.
\end{equation}
The quoted errors are the intercepts for FWHM. As can be seen in
figure~\ref{fig:richmap}, for the SE population of galaxies,
$\lambda(z)$ steadily increases up to $z=0.42$ to vanish at $z=0.7$
(not represented). For these redshifts, the choice of the color filter
$g-r$ is no longer accurately tracing red-sequence galaxies, but we can,
however, conclude that most of these objects do not belong to the
population east from the SZ center.

\begin{figure}
  \begin{center}
    \includegraphics[width=\onefigw]{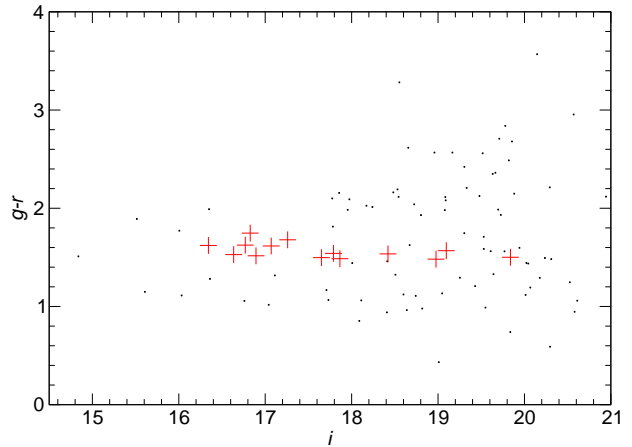}
    \caption{Color-magnitude diagram for the all objects located
      within a $R_\mathrm{c}(\lambda_{\max}) \simeq 1.1\,\Mpc$ radius
      around the cluster center (see the text and
      Fig.~\ref{fig:richmap}). Red crosses are for galaxies identified
      as cluster members, while the black dots are for all the other
      objects.}
\label{fig:cmd}
  \end{center}
\end{figure}

Finally, in figure~\ref{fig:cmd}, we have represented the
color-magnitude diagram for all objects found within a
$R_\mathrm{c}(\lambda_{\max})$ radius around the location $(101.31,
59.48)$, corresponding to the east population of
figure~\ref{fig:richmap}. The red crosses represent red-sequence
galaxies, which have been identified as belonging to the cluster with
a probability greater than $0.5$. The scatter of half a magnitude in
$g-r$ for bright objects ($i <17$) might be the result of the
aforementioned systematics in the zero-points. As already discussed,
they may bias $\lambda$ to lower values by pulling galaxies off the
red sequence. For fainter objects, the dispersion in $g-r$ shows a
clear degradation, which is expected due to the loss of signal,
especially toward the limiting magnitude $i\gtrsim 20$.

\section{Discussion}
\label{sec:discuss}

\subsection{Comparison with Pan-STARRS1}
\label{sec:panstarrs}

The SZ source under scrutiny belongs to the second Planck SZ catalog,
and it does not appear in the Pan-STARRS follow-up
paper~\citep{Liu:2014rza}. However, the recent data release of the
Pan-STARRS1 (PS1) survey~\citep{2016arXiv161205560C} covers the
location of {\aapeVa}. As can be checked on the PS1 public
archive\footnote{\href{http://ps1images.stsci.edu/cgi-bin/ps1cutouts?pos=101.23488+59.450954}{https://confluence.stsci.edu/display/PANSTARRS}},
the $grizy$ stack images encompassing the SZ location show significant
background subtraction and saturation artifacts coming from the
central bright star. The depth of the $griz$ channels also appears to
be smaller than the one we have obtained from the $14$ hours of
exposure (see Fig.~\ref{fig:griz3}). This is not very surprising, as
the small aperture of the $0.6\,\meter$ telescope renders imaging less
sensitive to saturation\footnote{Two-colors fits available at this
  \href{http://cp3.irmp.ucl.ac.be/~ringeval/upload/aape5/twocolors}{url}}.
Concerning the limiting magnitudes, as explained in
\citet{2016arXiv161205560C}, although the Pan-STARRS telescope
collects $10$ times more light than the T62, the exposure time on each
patch is relatively small. For the plates encompassing {\aapeVa}, this
is about $10$ minutes for the $grz$ channels, and $20$ minutes for
$i$. Nevertheless, the objects we have identified as galaxies are also
present in the PS1 redder colors, while being hardly visible in $g$
only. More interestingly, and although we have not attempted a
quantitative comparison, the other southeast galaxy overdensity is
present and more easily seen owing to the larger field of view
accessible in the PS1.

\subsection{Other redshift estimates}

We have also attempted to estimate the photometric redshift associated
with the $griz_3$ catalog by using a simple linear color fit over
the $g$, $r$, $i$ and $z$ magnitudes. The best fit has been determined
over $100\,000$ galaxies of the SDSS DR9 catalog by iterative
rejections of all objects remaining at more than $3 \sigma$ from the
best fit. This rejection is necessary as a brute force linear color
fit over all SDSS DR9 galaxies ends up being reasonably tracking the
photometric redshift only within the limited range $0.1 \lesssim
\zphot \lesssim 0.35$. Applying the best fit to the $griz_3$ catalog
yields a photometric redshift distribution, again centered around a
redshift $0.3$, but with a much larger spreading than the one
represented in Fig.~\ref{fig:griz3}. We find $\zphot=0.35 \pm 0.16$,
the first value being again the mode and the second one stands for the
absolute median deviation.

Finally, we have tested an alternative redshift estimate by ignoring
data taken with the $z'$ filter, as it shows some higher photometric
errors than in the $g'$, $r'$, and $i'$ bands for faint objects (see
Sect.~\ref{sec:reduc}). Following in all points the method presented
in Sect.~\ref{sec:results}, but for the $gri$ magnitudes only, one
finds, within the $3\arcmin$-radius region, $\zphot=0.27 \pm 0.06$ for
the mode and absolute median deviation. This is again compatible with
the photometric estimate in Eq.~\eqref{eq:zphot3}.

\subsection{Instrumental tests on confirmed clusters}

As mentioned in the introduction, instrumental tests have been
undertaken during the 2015 mission to assess the feasibility of
resolving individual galaxies in distant clusters with small
telescopes. Successful identification of individual galaxies has been
obtained for all the observed clusters, and we briefly report below our
results for each of them.

\begin{figure}
\begin{center}
  \includegraphics[width=\onefigw]{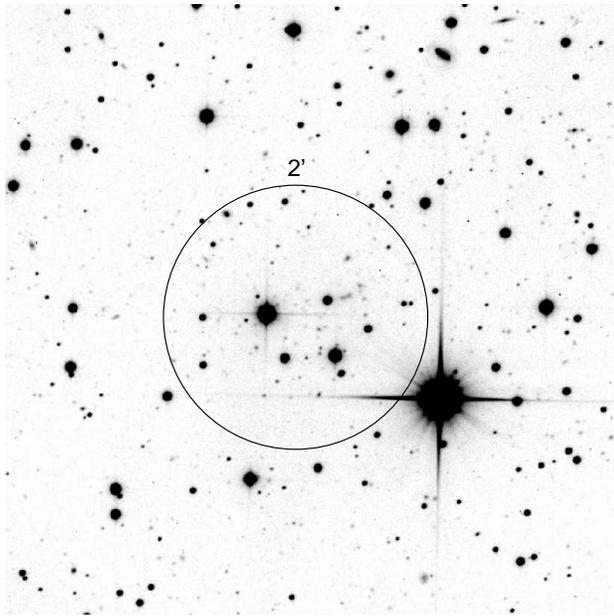}
  \caption{Low-signal instrumental test for the $0.6\,\meter$
    telescope during the 2015 mission. The image corresponds to a
    total of $1.5$ hours of luminance exposure. Many galaxies of
    {\aapeIIIcbis}, located at redshift $z=0.584$, are individually
    resolved, but the elliptic shape of only a few can be guessed.}
\label{fig:farmacs}
\end{center}
\end{figure}

\begin{description}

\item[\aapeIIIc], referred to as {\aapeIIIcbis}, is a massive and
  distant cluster located at $z=0.584$. It has been used as a
  low-signal test of the T62 and imaged in luminance for a total
  exposure time of $1.5$ hours. As can be seen in
  Fig.~\ref{fig:farmacs}, individual galaxies are resolved, but the
  elliptical shape of only a few can be guessed. Overdensity in galaxy
  counts is obvious, but occupies a relatively small angular size,
  about $2\arcmin$.\\

\item[\aapeIIId], also known as {\aapeIIIdbis}, is reported in the
  {\planck} catalog with a X-ray redshift of
  $\zx=\zaapeIIId$~\citep{2003ApJ...594..154M}. The elliptic shape of
  the most massive galaxies is resolved after less than $1$ hour of
  luminance exposure.\\

\item[\aapeIIIe], also {\aapeIIIebis}, at a redshift of
  $z=\zaapeIIIe$~\citep{2001ApJ...553..668E}. Galaxies are
  individually resolved after $1.6$ hours of luminance exposure, but
  elliptic shape can only be inferred for the largest ones.\\

\item[\aapeIIIa] has a X-ray counterpart known as
  {\aapeIIIabis}, which is located at a redshift
  $\zspec=0.280$~\citep{Barr:2005gi}. This cluster has been observed
  with the $0.5\,\meter$ telescope and four non-standard
  interferential color filters available at that time, centered over
  the red, green, blue and near infrared wavelengths, with about $2.5$
  hours of exposure per filter. Calibration over SDSS galaxies was
  used to determine a transformation matrix between these colors and
  the SDSS $ugri$ ${\modelmag}$, and a linear color-redshift fit
  gives $\zphot=0.30 \pm 0.1$ (mode and absolute median
  deviation).\\

\item[\aapeIIIb], a close cluster belonging to the
  \citet{1958ApJS....3..211A} catalog and known as {\aapeIIIbbis},
  at a redshift of $z=0.0739$. It has been observed in luminance with
  the T62 and for a total exposure time of $4.5$ hours. The shape of
  galaxies is well resolved, and their internal structure, such as
  spiral arms, can be observed for many of them. However, no obvious
  galaxy overdensity can be inferred, as the $14\arcmin$ field of view
  is too small for encompassing all the cluster's galaxies. We have
  also observed the same field of view with the $0.5\,\meter$
  telescope in three non-standard colors (blue, green, and near
  infrared) up to $1.5$ hours per filter. A linear color-redshift fit
  gives $\zphot=0.09 \pm 0.08$ for the mode and the absolute median
  deviation.

\end{description}

\section{Conclusions}
\label{sec:conclu}

This article presents the results of optical follow-up searches of
{\planck} SZ sources with non-professional $0.5\,\meter$ and
$0.6\,\meter$ telescopes. Our main result is the confirmation of the
cluster candidate {\aapeVa} by observing more than $100$ galaxies
within a $3\arcmin$-radius region around the SZ coordinates and by
estimating their photometric redshift at $\zphot=0.29 \pm 0.08$. We
have also estimated the optical richness, $\lambda$, around this
location. Although systematics in the photometric zero-points are
expected to bias $\lambda$ to lower values, its dependence in redshift
allows us to narrow down the cluster's redshift to
$z_\lambda=0.30{\,}^{+0.03}_{-0.05}$. Imaging also reveals a
contiguous population of galaxies extending a few arcminutes
southeast from the SZ region. However, their optical richness
suggests that they do not belong to the cluster, and are loosely
concentrated around $z \simeq 0.42$.

The robustness of these results have been tested against the choice of
the region in which galaxies are counted, the color-redshift
relationship, and by removing the $z'$ color data. Moreover,
photometric redshift estimates of two already confirmed clusters of
the {\pszcat} catalog, made with the same telescopes, are compatible
with their actual redshifts.

The potential of individual sub-meter instruments in the optical
confirmation of SZ sources and the measurement of their redshift is
obviously limited with respect to what can be achieved with much
larger telescopes. However, compared with automated acquisition
campaigns, their advantage is the possibility to acquire sensibly
deeper data by accumulating longer exposure time over one target (see
Sect.~\ref{sec:panstarrs}). The present results suggest that observing
galaxy overdensity and estimating redshift are achievable in a
redshift range $0.1$ to $0.6$, albeit with relatively large
uncertainties for the redshift, up to $40\%$ for the closest
clusters. In addition, there are potentially hundreds of sub-meter
class telescopes available around the world, mostly within the amateur
astronomer community. With even a fraction of them, dedicated searches
of SZ sources could be collectively envisaged, in a way similar to
what the Galaxy Zoo project is
providing~\citep{2008MNRAS.389.1179L}. A case-by-case tuning of the
acquisition parameters (unitary exposure time in particular) can be
operated to account for the presence of bright objects in the vicinity
of the faint galaxies to be detected, which is more complicated to
achieve in the case of an automatic survey. Such a collective campaign
would certainly bring complementarity to automated surveys. At the
very least, such a project could be used to provide a first
information on the redshift before starting more accurate
investigations with professional telescopes.

\section*{Acknowledgments}

We would like to warmly thank the Astroqueyras Association for having
allocated us the time and resources needed to carry on this
project. In particular, the acquisition of the set of Sloan filters
was greatly appreciated. It is a pleasure to thank E.~Bertin,
H.~J.~McCracken, and the two anonymous referees for precious
advice. We also acknowledge usage of the SDSS-III data, whose funding
and participating institutions can be found at
\url{http://www.sdss3.org/}. The optical richness code has been
developed based on the one publicly provided by R.~H.~Wechsler at
\url{http://risa.stanford.edu/redmapper}. We have made use of the NED
database, operated by the Jet Propulsion Laboratory, the California
Institute of Technology, under NASA contract, and the SIMBAD database,
operated at the Centre de Donn\'ees Astronomique de Strasbourg,
France~\citep{2000A&AS..143....9W}. Preprocessing of raw images has
been made with IRAF, distributed by the National Optical Astronomy
Observatories. Catalog manipulations have been made using
\href{http://ds9.si.edu}{SAOImage DS9}.

\bibliographystyle{aasjournal}
\bibliography{references}

\end{document}